\documentclass{article}
\usepackage{amsmath}
\usepackage{amsfonts}
\usepackage{amssymb}

\begin{document}

\title{On extreme Bosonic linear channels}
\author{A.S.\ Holevo \\
{\small Steklov Mathematical Institute,} {\small Gubkina 8, 119991
Moscow, Russia} }
\date{}
\maketitle

\begin{abstract}
The set of all channels with fixed input and output is convex. We first give
a convenient formulation of necessary and sufficient condition for a channel
to be extreme point of this set in terms of complementary channel, a notion
of big importance in quantum information theory. This formulation is based
on the general approach to extremality of completely positive maps in an
operator algebra due to Arveson. We then apply this formulation to prove the
main result of this note: under certain nondegeneracy conditions, purity of
the environment is necessary and sufficient for extremality of Bosonic
linear (quasi-free) channel. It follows that Gaussian channel between
finite-mode Bosonic systems is extreme if and only if it has minimal noise.
\end{abstract}

\section{Extremality in terms of complementary channels}

In what follows $\mathcal{H}$ (possibly with indices) denotes a separable
Hilbert space, $\mathfrak{T}(\mathcal{H})$ -- the Banach space of
trace-class operators and $\mathfrak{B}(\mathcal{H})=\mathfrak{T}(\mathcal{H}
)^{\ast }$-- the algebra of all bounded operators in $\mathcal{H}$. Let $A,B$
be two quantum systems with the Hilbert spaces $\mathcal{H}_{A},\mathcal{H}
_{B},$ which we call the input and the output systems. In this paper we call
by \textit{channel} a normal, unital, completely positive map $\Phi :
\mathfrak{B}(\mathcal{H}_{B})$ $\longrightarrow \mathfrak{B}(\mathcal{H}
_{A}).$ There is a unique linear trace-preserving map $\Phi _{\ast }:
\mathfrak{T}(\mathcal{H}_{A})$ $\longrightarrow \mathfrak{T}(\mathcal{H}
_{B}) $ \ such that $\Phi =\left( \Phi _{\ast }\right) ^{\ast },$ which maps
density operators (states) in $\mathcal{H}_{A}$ into density operators in $
\mathcal{H} _{B}.$ In physical terms we work in the Heisenberg picture,
while $\Phi _{\ast }$ is the channel in the Schr\"{o}dinger picture.

The set of all channels with input $A$ and output $B$ is convex. We first
give a convenient formulation of necessary and sufficient condition for a
channel to be extreme point of this set in terms of complementary channel, a
notion of big importance in quantum information theory \cite{ds,h,mr}. This
formulation is based on the general approach to extremality of completely
positive maps of a $C^*$-algebra due to Arveson \cite{ar} (see also
Appendix). We then apply this formulation to prove the main result of this
note: under certain nondegeneracy conditions, purity of the environment is
necessary and sufficient for extremality of Bosonic linear (quasi-free)
channel. It follows that a Gaussian channel between finite-mode Bosonic
systems is extreme if and only if it has minimal noise. For channels in one
Bosonic mode this was conjectured by Ivan, Sabapathy and Simon \cite{ivan}
basing on consideration of finite-dimensional criterion of Choi \cite{choi}.
Finding the proof for this statement was the initial motivation of the
present work.

Given three quantum systems $A,B,E$ with the spaces $\mathcal{H}_{A},
\mathcal{H}_{B},\mathcal{H}_{E}$ and an isometric operator $V:\mathcal{H}
_{A}\rightarrow \mathcal{H}_{B}\otimes \mathcal{H}_{E}$, the relations
\begin{eqnarray}
\Phi \lbrack X] &=&V^{\ast }(X\otimes I_{E})V,\qquad X\in \mathfrak{B}(
\mathcal{H}_{B}),  \label{compl} \\
\quad \tilde{\Phi}[Y] &=&V^{\ast }(I_{B}\otimes Y)V,\qquad Y\in \mathfrak{B}
( \mathcal{H}_{E})  \label{compl1}
\end{eqnarray}
define two channels $\Phi :\mathfrak{B}\left( \mathcal{H}_{B}\right)
\rightarrow \mathfrak{B}\left( \mathcal{H}_{A}\right) ,$ $\tilde{\Phi}:
\mathfrak{B}\left( \mathcal{H}_{E}\right) \rightarrow \mathfrak{B}\left(
\mathcal{H}_{A}\right) ,$ which are called mutually \textit{complementary}.

The Stinespring dilation theorem implies that for given a channel $\Phi $
the representation (\ref{compl}) and hence a complementary channel $\tilde{
\Phi}$ given by (\ref{compl1}) always exists. The representation (\ref{compl}) is \textit{minimal} if the subspace
\begin{equation*}
\mathcal{M}=\{(X\otimes I_{E})V\psi :\psi \in \mathcal{H}_{A},X\in
\mathfrak{ \ \ \ \ B}(\mathcal{H}_{B})\}\subset
\mathcal{H}_{B}\otimes \mathcal{H}_{E}
\end{equation*}
is dense in $\mathcal{H}_{B}\otimes \mathcal{H}_{E}.$ Let
\begin{equation}
\Phi \lbrack X]=V^{\prime \ast }(X\otimes I_{E^{\prime }})V^{\prime },\qquad
X\in \mathfrak{B}(\mathcal{H}_{B})  \label{sties1}
\end{equation}
be another representation for $\Phi ,$ then there exists an isometric
operator $W$ from $\mathcal{H}_{E}$ into $\mathcal{H}_{E^{\prime }}$ such
that
\begin{equation}
(I_{B}\otimes W)V=V^{\prime },  \label{WVtildeV}
\end{equation}
so that the new complementary channel is $\tilde{\Phi}^{\prime }[Y^{\prime
}]=\tilde{\Phi}[W^{\ast }Y^{\prime }W].$ In particular, if the
representation (\ref{sties1}) is also minimal, the operator $W$ maps $
\mathcal{H}_{E}$ onto $\mathcal{H}_{E^{\prime }},$ so that the minimal
representation and the corresponding complementary channel are unique up to
the unitary equivalence. In this case the complementary channel is also
called minimal.

Let $Y$ be an operator in $\mathcal{H}_{E}$ such that $0\leq Y\leq I_{E},$
then the relation
\begin{equation*}
\Phi _{Y}[X]=V^{\ast }(X\otimes Y)V,\qquad X\in \mathfrak{B}(\mathcal{H}
_{B}),
\end{equation*}
apparently defines a completely positive map satisfying $\Phi _{Y}\leq \Phi
. $ The operator-algebraic version of the Radon-Nikodym theorem established
by Arveson \cite{ar}, Theorem 1.4.2, implies: assuming that the
representation (\ref{compl}) is minimal, this relation sets one-to-one
affine correspondence between the order intervals $[0,I_{E}]$ and $[0,\Phi
]. $

\textbf{Proposition 1.} \textit{Channel $\Phi $ is extreme if and only if it
has a complementary channel $\tilde{\Phi}^{\prime }$ such that $\mathrm{Ker}
\tilde{\Phi}^{\prime }=0,$ or, equivalently, $\overline{\mathrm{Ran}\tilde{
\Phi}_{\ast }^{\prime }}=\mathfrak{T}(\mathcal{H}_{E}).$ Any such
complementary channel is minimal.}

This follows from Theorem 1.4.6 \cite{ar} (see Appendix) but we give
here a proof for completeness. Assume that $\mathrm{Ker}\tilde{\Phi}
^{\prime }=0,$ and let $\tilde{\Phi}$ be a complementary channel in
the minimal representation (\ref{compl}) for $\Phi ,$ so that
$\tilde{\Phi} ^{\prime }[Y^{\prime }]=\tilde{\Phi}[W^{\ast
}Y^{\prime }W]$ where $W^{\ast }W=I_{E}.$ Then $\tilde{\Phi}^{\prime
}[I_{E^{\prime }}-WW^{\ast }]=\tilde{ \Phi} [W^{\ast }\left(
I_{E^{\prime }}-WW^{\ast }\right) W]=0,$ therefore $ I_{E^{\prime
}}-WW^{\ast }=0,$ so that $W$ is unitary operator onto $
\mathcal{H}_{E^{\prime }}.$ Therefore we can assume
$\tilde{\Phi}^{\prime }= \tilde{\Phi}$ and
$\mathrm{Ker}\tilde{\Phi}=0.$ Let $\Phi =\frac{1}{2}\left( \Phi
_{1}+\Phi _{2}\right) ,$ where $\Phi _{j}$ are channels. Then by
Arveson's Theorem, $\Phi _{j}=\Phi _{Y_{j}};j=1,2,$ where $0\leq
Y_{j}\leq 2I_{E}.$ Since $\Phi _{j}$ are unital, $V^{\ast
}(I_{B}\otimes Y_{j})V=V^{\ast }(I_{B}\otimes I_{E})V,$ whence
$V^{\ast }(I_{B}\otimes \left( I_{E}-Y_{j}\right) )V=0.$ But this
means that $\tilde{\Phi} [I_{E}-Y_{j}]=0,$ hence $I_{E}-Y_{j}=0$.
Therefore $\Phi _{j}=\Phi ,$ so $ \Phi $ is extreme channel.

Conversely, let $\Phi $ be extreme channel and let us show that $\mathrm{Ker}
\tilde{\Phi}=0$ for the minimal complementary channel. Let $\tilde{\Phi}
[Y]=0.$ Without loss of generality we can assume that $Y=Y^{\ast }$ and $
\left\Vert Y\right\Vert \leq 1.$ Define $\Phi _{\pm }[X]=V^{\ast }(X\otimes
\left( I_{B}\pm Y\right) )V,$ for the minimal representation (\ref{compl})
of the channel $\Phi .$ Then $\Phi _{\pm }$ are normal completely positive
maps by Arveson's Theorem and $\Phi _{\pm }[I_{B}]=I_{A\text{ }}$by the
assumption. Since $\Phi $ is extreme, $\Phi =\Phi _{\pm },$ implying $
V^{\ast }(X\otimes Y)V=0$ for all $X,Y.$ It follows that $\left\langle \psi
_{2}|V^{\ast }(X_{2}^{\ast }\otimes I_{E})(I_{B}\otimes Y)(X_{1}\otimes
I_{E})V|\psi _{1}\right\rangle \equiv 0,$ hence the bilinear form of the
operator $I_{B}\otimes Y$ vanishes on $\mathcal{M}$, hence $Y=0$ by the
minimality of the representation.

\textbf{Remark.} Choosing an orthonormal basis $\left\{ e_{j}\right\} $ in $
\mathcal{H} _{E},$ introduce the operators $V_{j}:\mathcal{H}_{A}$ to $
\mathcal{H}_{B}$, defined by $V_{j}={\langle }e_{j}|\,V$ i.e.
\begin{equation}
{\langle }\phi |\,V_{j}\psi {\rangle }={\langle }\phi \otimes
e_{j}|\,V\psi { \rangle },\quad \phi \in \mathcal{H}_{B},\psi \in
\mathcal{H}_{A}, \label{braA}
\end{equation}
and let $y_{jk}={\langle }e_{j}|\,Ye_{k}{\rangle }$ be the matrix of the
operator $Y.$ Then the above result amounts to the following: \textit{\
channel $\Phi $ is extreme if and only if it has a representation}
\begin{equation}
\Phi [X]=\sum_{k}V_{k}^{\ast }XV_{k},\qquad X\in \mathfrak{B}( \mathcal{H}
_{B}),  \label{kr}
\end{equation}
\textit{where the system $\left\{ V_{j}^{\ast }V_{k}\right\} $ is strongly
independent in the sense that $\sum_{jk}y_{jk}V_{j}^{\ast }V_{k}=0$ (strong
operator convergence) for a matrix $\left[ y_{jk}\right] $ of bounded
operator implies $y_{jk}\equiv 0$.} This result contained in \cite{ts}
generalizes Choi's criterion for finite dimensional case \cite{choi}.
However for our purposes the formulation in terms of the complementary
channel turns out to be more convenient.

\section{Extremality of Linear Bosonic Channels}

In what follows we consider Bosonic system with $s$ modes described by
irreducible Weyl-Segal system
\begin{equation}
W(z)=\exp i\,(Rz),  \label{1.1}
\end{equation}
in a Hilbert space $\mathcal{H}$, where
\begin{equation}
Rz=\sum_{j=1}^{s}(x_{j}q_{j}+y_{j}p_{j}),
\end{equation}
so that $R=[q_{1},p_{1},\dots ,q_{s},p_{s}]$ is the row vector of the
\textit{canonical observables } and $z=[x_{1},y_{1},\dots
,x_{s},y_{s}]^{\top }$ is the column vector of real parameters. We refer to
\cite{ho1,dvv,cgh,ssa} for relevant definitions and results.

A real vector space $Z$ equipped with a nondegenerate antisymmetric
bilinear form $\Delta $ is called symplectic space. Particularly
important case is the standard symplectic space $Z=\mathbb{R}^{2s}$
equipped with the form $ z^{\top }\Delta z^{\prime },$ where
\begin{equation}
\Delta =\left[
\begin{array}{ccccc}
0 & -1 &  &  &  \\
1 & 0 &  &  &  \\
&  & \ddots &  &  \\
&  &  & 0 & -1 \\
&  &  & 1 & 0
\end{array}
\right]  \label{delta}
\end{equation}
is the matrix of commutators of the canonical observables, $[Rz,Rz^{\prime
}]=-iz^{\top }\Delta z^{\prime }I.$

The noncommutative Fourier transform of a trace class operator
$\tau$ in $ \mathcal{H}$ is defined as
\begin{equation*}
\phi _{\tau}(z)=\mathrm{Tr}\tau W(z).
\end{equation*}
The complex function $\phi _{\tau}(z)$ is
bounded and continuous on $Z.$ If $\rho $ is a density operator (state), $\phi _{\rho}$ is called its
characteristic function. Then $\phi _{\rho}(0)=1$. Operator $\tau$ is positive if and only if $\phi
_{\tau}(z)$ is $\Delta - $\textit{nonnegative definite}: all the matrices
with the elements
\begin{equation}
\phi _{\tau}(z_{r}-z_{s})\,\exp \frac{i}{2}\bigl(z_{r}^{\top }\Delta
z_{s} \bigr) ,  \label{pd}
\end{equation}
where $z_{1},\ldots ,z_{n}$ is an arbitrary finite subset of $Z$, are
nonnegative definite. 

Let $(Z_{A},\Delta _{A}),(Z_{B},\Delta _{B})$ be the symplectic spaces of
dimensionalities $2s_{A},2s_{B},$ which will describe the input and the
output of the channel (here $\Delta _{A},\Delta _{B}$ have the canonical
form (\ref{delta})) and let $W_{A}(z_{A}),W_{B}(z_{B})$ be the Weyl
operators in the Hilbert spaces $\mathcal{H}_{A},\mathcal{H}_{B}$ of the
corresponding Bosonic systems. Channel $\Phi $ transforming the Weyl
operators according to the rule
\begin{equation}
\Phi \lbrack W_{B}(z_{B})]=W(Kz_{B})f(z_{B}),  \label{linbos}
\end{equation}
where $K$ is a linear map between output and input symplectic spaces and $f$
is a complex continuous function such that $f(0)=1$, is called \textit{linear Bosonic} \cite{hol} or
quasi-free \cite{dvv}. Define the real skew-symmetric $2s_{B}\times 2s_{B}-$
matrix
\begin{equation*}
\Delta _{K}=\Delta _{B}-K^{\top }\Delta _{A}K.
\end{equation*}
The map (\ref{linbos}) is completely positive if and only if $f$ is $\Delta
_{K}-$nonnegative definite: all the matrices with the elements
\begin{equation}
f(z_{r}-z_{s})\,\exp \frac{i}{2}\bigl(z_{r}^{\top }\Delta _{K}z_{s}\bigr),
\label{twcp}
\end{equation}
where $z_{1},\ldots ,z_{n}$ is an arbitrary finite subset of $Z$, are
nonnegative definite \cite{dvv},\cite{hw}. In what follows we assume that
\textit{\ the real skew-symmetric $2s_{B}\times 2s_{B}-$matrix $\Delta
_{K}=\Delta _{B}-K^{\top }\Delta _{A}K$ is nondegenerate, i.e.}
\begin{equation}
\det \Delta _{K}\neq 0.  \label{nondeg}
\end{equation}

Under this condition there exist real nondegenerate $2s_{B}\times 2s_{B}-$
matrix $K_{D}$ such that
\begin{equation}
\quad K_{D}^{\top }\Delta _{D}K_{D}=\Delta _{K},  \label{iskom2}
\end{equation}
where $\Delta _{D}=\Delta _{B}.$ This is just the canonical form of the
nondegenerate skew-symmetric matrix $\Delta _{K}$. Comparing (\ref{twcp})
with (\ref{pd}) we find that there exists a state $\rho _{D}$ of the Bosonic
system in the space $\mathcal{H}_{D}$ corresponding to the standard
symplectic space $(Z_{D},\Delta _{D})\simeq (Z_{B},\Delta _{B})$ such that
\begin{equation}
f(z_{B})=\mathrm{Tr}\rho _{D}W_{D}(K_{D}z_{B})=\phi _{\rho _{D}}(K_{D}z_{B})
\label{iskom}
\end{equation}
i.e.
\begin{equation}
\Phi \lbrack W_{B}(z_{B})]=W(Kz_{B})\phi _{\rho _{D}}(K_{D}z_{B}).
\label{qfc}
\end{equation}
The relation (\ref{iskom2}) implies that
\begin{equation}
\Delta _{B}=K^{\top }\Delta _{A}K+K_{D}^{\top }\Delta _{D}K_{D}.  \label{com}
\end{equation}

We will make use of the unitary dilation of the channel $\Phi $ from
\cite {cgh}. Consider the composite Bosonic system $AD=BE$ with the
Hilbert space $ \mathcal{H}_{A}\otimes \mathcal{H}_{D}\simeq
\mathcal{H}_{B}\otimes \mathcal{ H}_{E}$ corresponding to the
symplectic space $Z=Z_{A}\oplus Z_{D}=Z_{B}\oplus Z_{E},$ where
$(Z_{E},\Delta _{E})\simeq (Z_{A},\Delta _{A})$. Thus
$[R_{A}\,R_{D}]=[R_{B}\,R_{E}]$ describe different splits of the set
of canonical observables for the composite system. The channel $\Phi
$ is described by the linear input-output relation (preserving the
commutators as follows from (\ref{com})
\begin{equation}
R_{B}^{\prime }=R_{A}K+R_{D}K_{D}  \label{ior}
\end{equation}
where the system $D$ is in the state $\rho _{D}$ (for simplicity of
notations we write $R_{A},\dots $ instead of $R_{A}\otimes I_{D},\dots $).
It is shown that the commutator-preserving relation (\ref{ior}) can be
complemented to the full linear canonical transformation by putting
\begin{equation}
R_{E}^{\prime }=R_{A}L+R_{D}L_{D},  \label{iocomp}
\end{equation}
where $\left( 2s_{A}\right) \times \left( 2s_{A}\right) -$ matrix
$L$ and $ \left( 2s_{B}\right) \times \left( 2s_{A}\right) -$ matrix
$L_{D}$ are such that the $2\left( s_{A}+s_{B}\right) \times 2\left(
s_{A}+s_{B}\right) -$ matrix
\begin{equation}
T=\left[
\begin{array}{cc}
K & L \\
K_{D} & L_{D}
\end{array}
\right]  \label{bltr}
\end{equation}
is symplectic, i.e. satisfies
\begin{equation*}
T^{\top }\left[
\begin{array}{cc}
\Delta _{A} & 0 \\
0 & \Delta _{D}
\end{array}
\right] T=\left[
\begin{array}{cc}
\Delta _{B} & 0 \\
0 & \Delta _{E}
\end{array}
\right]
\end{equation*}
and $\Delta _{E}=\Delta _{A}.$

\textbf{Lemma 1.} \textit{Under the condition (\ref{nondeg}), $\det
L\neq 0.$ }

\textit{Proof.} The fact that $T$ is symplectic implies, in addition
to (\ref {com}),
\begin{eqnarray*}
\quad 0 &=&K^{\top }\Delta _{A}L+K_{D}^{\top }\Delta _{D}L_{D}, \\
\Delta _{E} &=&L^{\top }\Delta _{A}L+L_{D}^{\top }\Delta _{D}L_{D}.
\end{eqnarray*}
Taking into account that $\det K_{D}\neq 0,$ the first equation implies
\begin{equation*}
L_{D}=-\left( K_{D}^{\top }\Delta _{D}\right) ^{-1}K^{\top }\Delta _{A}L.
\end{equation*}
Substituting into the second equation gives $\Delta _{D}=L^{\top }ML,$ where
$M=\Delta _{A}+\Delta _{A}K\left( K_{D}\Delta _{D}\right) ^{-1}\Delta
_{D}\left( K_{D}^{\top }\Delta _{D}\right) ^{-1}K^{\top }\Delta _{A}.$
Therefore $\left( \det L\right) ^{2}\det M=1,$ whence $\det L\neq 0.\square $

Denote by the $U_{T}$ the unitary operator in $\mathcal{H}_{A}\otimes
\mathcal{H}_{D}\simeq \mathcal{H}_{B}\otimes \mathcal{H}_{E}$ implementing
the symplectic transformation $T$ so that
\begin{equation}
\lbrack R_{B}^{\prime }\,R_{E}^{\prime }]=U_{T}^{\ast
}[R_{B}\,R_{E}]U_{T}=[R_{A}\,R_{D}]T.  \label{deistvo}
\end{equation}
Then we have the unitary dilation
\begin{equation}
\Phi \lbrack W_{B}(z_{B})]=\mathrm{Tr}_{D}\left( I_{A}\otimes \rho
_{D}\right) U_{T}^{\ast }\left( W_{B}(z_{B})\otimes I_{E}\right) U_{T}.
\label{udi1}
\end{equation}
The \textit{weakly complementary} channel \cite{cgh} is then
\begin{equation*}
\tilde{\Phi}^{w}[W_{E}(z_{E})]=\mathrm{Tr}_{D}\left( I_{A}\otimes \rho
_{D}\right) U_{T}^{\ast }\left( I_{B}\otimes W_{E}(z_{E})\right) U_{T}.
\end{equation*}
The equation (\ref{iocomp}) is nothing but the input-output relation for the
weakly complementary channel which thus acts as
\begin{equation}
\tilde{\Phi}^{w}[W_{E}(z_{E})]=W_{A}(Lz_{E})\phi _{\rho _{D}}(L_{D}z_{E}).
\label{Gc}
\end{equation}
In the case of pure state $\rho _{D}=|\psi _{D}\rangle \langle \psi
_{D}|$ the relation (\ref{udi1}) amounts to the Stinespring
representation (\ref {compl}) for the channel $\Phi $ with the
isometry $V=U_{T}|\psi _{D}\rangle ,$ and the relation (\ref{udi1})
amounts to (\ref{compl1}) implying that $
\tilde{\Phi}^{w}=\tilde{\Phi}.$

Apparently if the channel $\Phi $ given by (\ref{qfc}) is extreme then $\rho
_{D}$ is a pure state (otherwise the spectral decomposition of $\rho _{D}$
would provide a nontrivial convex decomposition of $\Phi $). In the converse
direction we prove

\textbf{Theorem.} \textit{Assume that} $\rho _{D}$ \textit{is a pure
state with nonvanishing characteristic function } $\phi _{\rho
_{D}}$ \textit{ which is } $ L^{2}-$\textit{differentiable to the
order} $2s_{D}$. \textit{Then the channel $\Phi $ given by
(\ref{qfc}) is extreme.}

\textbf{Remark.} We conjecture that a similar result should hold without
assumption (\ref{nondeg}) for a Bosonic linear channels on the CCR-algebra.
In \cite{dvv} purity of $\rho _{D}$, in the situation where $K$ is
symplectic transformation or symplectic projection (so that $\det \Delta
_{K}=0$), was shown to be sufficient for extremality of a quasi-free map on
the CCR-algebra.

\textit{Proof.} We shall show that the complementary channel $\tilde{\Phi}$
satisfies the condition $\overline{\mathrm{Ran}\tilde{\Phi}_{\ast }}=
\mathfrak{T}(\mathcal{H}_{E})$ of the Proposition 1.

\textbf{Lemma 2.} \textit{Let }$\mathit{\Psi _{K,f}}$\textit{\ be a Bosonic
linear channel with the same input and output space $\mathcal{H}$,}
\begin{equation*}
\Psi_{K,f} \lbrack W(z)]=W(Kz)f(z),
\end{equation*}
\textit{\ where $K$ is a nondegenerate square matrix. Then the restriction
of $\Psi _{K,f}$ onto $\mathfrak{T}(\mathcal{H})$ coincides with $\left\vert
\det K\right\vert ^{-1}\left( \Psi _{\hat{K},\hat{f}}\right) _{\ast },$
where $\hat{K}=K^{-1},\hat{f}(z)=f(-K^{-1}z).$}

\textit{Proof.} We use the inversion formula for the noncommutative Fourier
transform \cite{ho1}: if $\tau$ is trace-class operator and $\phi _{\tau}(z)=\mathrm{Tr}
\tau W(z), $ then
\begin{equation*}
\tau=\frac{1}{\left( 2\pi \right) ^{s}}\int \phi _{\tau}(z)W(-z)d^{2s}z.
\end{equation*}
It follows that
\begin{eqnarray*}
\Psi _{K,f}[\tau] &=&\frac{1}{\left( 2\pi \right) ^{s}}\int \phi
_{\tau}(z)W(-Kz)f(-z)d^{2s}z \\
&=&\frac{1}{\left( 2\pi \right) ^{s}\left\vert \det K\right\vert }\int \phi
_{\tau}(K^{-1}z)W(-z)f(-K^{-1}z)d^{2s}z \\
&=&\frac{1}{\left( 2\pi \right) ^{s}\left\vert \det K\right\vert }\int \phi
_{\tau}(\hat{K}z)W(-z)\hat{f}(z)d^{2s}z.
\end{eqnarray*}
Then
\begin{eqnarray*}
\mathrm{Tr}\Psi _{K,f}[\tau]W(z) &=&\left\vert \det K\right\vert ^{-1}\phi
_{\tau}( \hat{K}z)\hat{f}(z) \\
&=&\left\vert \det K\right\vert ^{-1}\mathrm{Tr}\tau W(\hat{K}z)\hat{f}(z) \\
&=&\left\vert \det K\right\vert ^{-1}\mathrm{Tr}\tau\Psi _{\hat{K},\hat{f}
}(W(z))
\end{eqnarray*}
and the Lemma is proved.$\square $

\textbf{Lemma 3.} \textit{Under the condition of Lemma 2,
}$\overline{ \mathit{\mathrm{Ran}\tilde{\Phi}_{\ast
}}}=\mathfrak{\ T}(\mathcal{H}_E).$

\textit{Proof.} From Lemma 2 it follows that up to a positive factor
$\tilde{ \Phi}_{\ast }$ is itself Bosonic linear channel satisfying
the condition of Lemma 2. It is sufficient to prove that arbitrary
positive trace-class operator $\tau$ in $\mathcal{H}_E$ can be
approximated in the trace norm by operators of the form $\Phi _{\ast
}[\tau_{n}],\tau_{n}\in \mathfrak{T}( \mathcal{H}_A ).$ By the
Parceval identity for the noncommutative Fourier transform, the
function $\phi _{\sqrt{\tau}}(z)$ is square integrable. Denote by
$\mathcal{C}$ the class of infinitely differentiable functions with
finite support. Let $\left\{ \phi _{n}(z)\right\} \subset
\mathcal{C}$ be a sequence converging to $\phi _{\sqrt{\tau}}(z)$ in
$L^{2}(Z).$ Consider the operators
\begin{equation*}
\sigma_{n}=\frac{1}{\left( 2\pi \right) ^{s}}\int \phi _{n}(z)W(-z)d^{2s}z.
\end{equation*}
We can assume that $\phi _{n}(-z)=\overline{\phi _{n}(-z)}$ so that
$\sigma_{n}$ is Hermitean. By the Parceval identity for the
noncommutative Fourier transform, the Hilbert-Schmidt norms
$\left\Vert \sqrt{\tau}-\sigma_{n}\right\Vert _{2}\longrightarrow
0.$ By using the inequality $\left\Vert AB\right\Vert _{1}\leq
\left\Vert A\right\Vert _{2}\left\Vert B\right\Vert _{2}$ we
conclude $ \left\Vert \tau-\sigma_{n}^{2}\right\Vert
_{1}\longrightarrow 0.$ The function $\varphi
_{n}(z)=\mathrm{Tr}\sigma_{n}^{2}W(z)$ is twisted convolution of two
functions $\phi _{n}(z)$ and hence also belongs to $ \mathcal{C}$.
Then by change of variables
\begin{eqnarray*}
\sigma_{n}^{2}&=&\frac{1}{\left( 2\pi \right) ^{s}}\int \varphi
_{n}(z)W(-z)d^{2s}z \\
&=& \frac{1}{\left( 2\pi \right) ^{s}\left\vert \det L\right\vert }\int
\varphi _{n}(L^{-1}z)W(-L^{-1}z)d^{2s}z=\tilde{\Phi}_{\ast }[\tau_{n}],
\end{eqnarray*}
where $\tau_{n}=\frac{1}{\left( 2\pi \right) ^{s}}\int \frac{\varphi
_{n}(L^{-1}z)}{f(L^{-1}z)}W(-z)d^{2s}z$ and $f(z)$ is given by
(\ref{iskom}), so it does not vanish and is $ L^{2}-$differentiable
to the order $2s$ by the assumption of Theorem. Hence the function
$\frac{\varphi _{n}(L^{-1}z)}{ f(L^{-1}z)}$ is well defined,
finitely supported and also $ L^{2}-$differentiable to the order
$2s$. It remains to show that $\tau_{n}$ is a trace class operator.

\textbf{Lemma 4.} \textit{If $\phi _{\tau }$  has finite support and
is }$ L^{2}-$\textit{differentiable of the order} $2s$
\textit{\ then $\tau \in
\mathfrak{T}(\mathcal{H}).$}

\textit{Proof.} By using the formula (see \cite{ho1}, Lemma V.4.2),
\begin{equation*}
\phi _{\tau (Rw)}(z)=\left[ -\frac{1}{2}z^{\top }\Delta w-\nabla _{w}\right]
\phi _{\tau }(z),\quad w,z\in Z,
\end{equation*}
we see that  $\phi _{\tau (Rw)^{2s}}$ is square integrable. It
follows that $ \tau (Rw)^{2s}$ extends to a Hilbert-Schmidt
operator, and similarly the operator $\sigma =\tau \left(
2N_{1}+1\right) \dots \left( 2N_{s}+1\right) ,$ where
$2N_{j}+1=\left( Re_{j}\right) ^{2}+\left( Rh_{j}\right)
^{2}=q_{j}^{2}+p_{j}^{2}\ $for a symplectic basis $\left\{
e_{j},h_{j}\right\} _{j=1,\dots ,s}$ in $Z.$ Here $N_{j}$ is the
number operator of $j-$th mode which is selfadjoint with the
eigenvalues $ n_{j}=0,1,\dots ,$ and the operators $N_{1},\dots
,N_{s}\ $commute. Therefore $\sigma ^{\ast }\sigma $ is a positive
trace class operator. From this we conclude
\begin{equation*}
\mathrm{Tr}\sigma ^{\ast }\sigma =\sum\limits_{n_{1},\dots n_{s}}\left(
2n_{1}+1\right) ^{2}\dots \left( 2n_{s}+1\right) ^{2}\langle n_{1},\dots
n_{s}|\tau ^{\ast }\tau |n_{1},\dots n_{s}\rangle <\infty ,
\end{equation*}
where $\left\{ |n_{1},\dots n_{s}\rangle \right\} $ is the orthonormal basis
of common eigenvectors of operators $N_{1},\dots ,N_{s}.$ By using
Cauchy-Schwarz inequality and the inequality $\langle \psi |\,|\tau |\,|\psi
\rangle ^{2}\leq \langle \psi |\tau ^{\ast }\tau |\psi \rangle $ for a unit
vector $\psi ,$ we have
\begin{eqnarray*}
\left( \mathrm{Tr}|\tau |\right) ^{2} &=&\left( \sum\limits_{n_{1},\dots
n_{s}}\langle n_{1},\dots n_{s}|\,|\tau |\,|n_{1},\dots n_{s}\rangle \right)
^{2} \\
&\leq &\sum\limits_{n_{1},\dots n_{s}}\left( 2n_{1}+1\right)
^{2}\dots \left( 2n_{s}+1\right) ^{2}\langle n_{1},\dots n_{s}|\tau
^{\ast }\tau |n_{1},\dots n_{s}\rangle
\\&\cdot & \sum\limits_{n_{1},\dots n_{s}}\left( 2n_{1}+1\right)
^{-2}\dots \left( 2n_{s}+1\right) ^{-2}<\infty ,
\end{eqnarray*}

Thus Lemma 4 and hence Lemma 3 are proved. Applying the Proposition 1 proves
the Theorem.$\square $

\section{\protect\bigskip The case of Gaussian channels}

The density operator (state) $\rho $ is called \textit{\ Gaussian}, if its
characteristic function $\phi _{\rho }(z)=\mathrm{Tr}\rho W(z)$ has the form
\begin{equation}
\phi _{\rho }(z)=\exp \left( il^{\top }z-\frac{1}{2}z^{T}\alpha z\right) ,
\label{GaussianState}
\end{equation}
where $\alpha $ is a real symmetric $(2s)\times (2s)$-matrix, called
covariance (or correlation) matrix of $\rho $. The necessary and sufficient
condition for $\alpha $ to be a covariance matrix is the inequality
\begin{equation}
\alpha \geq \frac{i}{2}\Delta ,  \label{n-s condition}
\end{equation}
where both parts are considered as complex Hermitian matrices. This is
equivalent to the $\Delta -$nonnegative definiteness of $\phi _{\rho }(z).$

\textbf{Proposition 2.} \textit{Gaussian state $\rho $ is pure if and only
one of the following equivalent conditions holds:}

\begin{enumerate}
\item \textit{$\alpha $ is a minimal (in the sense of partial ordering of
real symmetric matrices) solution of the inequality (\ref{n-s condition});}

\item \textit{the symplectic eigenvalues of the matrix $\alpha $ are all
equal to their minimal possible value $\frac{1}{2};$}

\item $\mathrm{Rank}\left( \alpha -\frac{i}{2}\Delta \right) =s;$

\item $\alpha +\frac{1}{4}\Delta \alpha ^{-1}\Delta =0$;

\item \textit{$\alpha =-\frac{1}{2}\Delta J$, where $J$ is an operator of
complex structure in $(Z,\Delta )$.}
\end{enumerate}

This statement is a collection of results scattered in literature; its proof
is essentially based on Williamson's canonical form of the matrix $\alpha $,
see \cite{ho1,lin,ssa}.

\bigskip \textit{Bosonic Gaussian channel} $\Phi =\Phi _{K,l,\mu }$ is
Bosonic linear channel with Gaussian function $f(z),$ namely
\begin{equation}
\Phi _{K,l,\mu }[W_{B}(z_{B})]=W(Kz_{B})\exp \left[ il^{\top }z_{B}-\frac{1}{
2}z_{B}^{\top }\mu z_{B}\right] ,  \label{bosgaus}
\end{equation}
Here $K$ is real $2s_{A}\times 2s_{B}-$matrix and $\mu $ is real
symmetric $ 2s_{B}\times 2s_{B}-$matrix satisfying the inequality
\begin{equation}
\mu \geq \frac{i}{2}\left[ \Delta _{B}-K^{\top }\Delta _{A}K\right] ,
\label{nis}
\end{equation}
which is necessary and sufficient condition for complete positivity. The
channel (\ref{bosgaus}) has \textit{minimal noise} if $\mu $ is a minimal
solution of this inequality \cite{lin,cgh}.

Assuming the condition (\ref{nondeg}), let $K_{D}$ be a solution of
(\ref {iskom2}), then $\alpha _{D}=\left( K_{D}^{\top }\right)
^{-1}\mu \left( K_{D}\right) ^{-1}\ $is real symmetric $2s_{B}\times
2s_{B}-$matrix such that $\alpha _{D}\geq \frac{i}{2}\Delta _{D}.$
Further, from minimality of $ \mu $ it follows that $\alpha _{D}$ is
a minimal solution of the inequality $ \alpha _{D}\geq
\frac{i}{2}\Delta _{D},$ and as such it is the covariance matrix of
a pure centered $(l=0)$ Gaussian state $\rho _{D}=|\psi _{D}\rangle
\langle \psi _{D}|$ of the Bosonic system in the space
$\mathcal{H}_{D}$ corresponding to the standard symplectic space
$(Z_{D},\Delta _{D})\simeq (Z_{B},\Delta _{B}).$ Applying
Theorem 1 to the case of Gaussian $\rho _{D}$ we
obtain

\textbf{Corollary.} \textit{Bosonic Gaussian channel is extreme if and only
if it has the minimal noise.}

Lemma 2 in the case of Gaussian channels amounts to the statement:

\textit{Let $\Phi _{K,l,\mu }$ be the Gaussian channel with the same input
and output space, where $K$ is a nondegenerate square matrix. Then the
restriction of $\Phi _{K,l,\mu }$ onto $\mathfrak{T}( \mathcal{H})$
coincides with $\left\vert \det K\right\vert ^{-1}\left( \Phi _{\hat{K},\hat{
l},\hat{\mu}}\right) _{\ast },$ where}
\begin{equation*}
\left( \hat{K},\hat{l},\hat{\mu}\right) =\left( K^{-1},-\left( K^{-1}\right)
^{\top }l,\left( K^{-1}\right) ^{\top }\mu K^{-1}\right) .
\end{equation*}

This generalizes the duality observed for the one-mode Gaussian channels in
the canonical form in \cite{ivan}.

\section{Appendix}

\bigskip Let $\mathfrak{B}$ be a $C^{\ast }-$algebra and $\Phi $ a
completely positive map from $\mathfrak{B}$ to $\mathfrak{B}(\mathcal{H}),$
where $\mathcal{H}$ is a separable Hilbert space. Let
\begin{equation*}
\Phi \lbrack X]=V^{\ast }\pi (X)V,\quad X\in \mathfrak{B},
\end{equation*}
be a minimal Stinespring representation for $\Phi ,$ where $V$ is a
bounded operator from $\mathcal{H}$ to $\mathcal{K}$, another
separable Hilbert space, and $\pi $ is a representation of
$\mathfrak{B}$ on $\mathcal{K}$. Let $\mathfrak{E}=\pi
(\mathfrak{B})^{\prime }$ be the commutant of the algebra $\pi
(\mathfrak{B})$ in $\mathcal{K}$. Theorem 1.4.6 in \cite{ar} says
that $\Phi $ \textit{is an extreme point of the convex set of
completely positive maps }$\Psi $\textit{\ normalized so that }$\Psi
\lbrack I]= N\equiv V^{\ast }V$\textit{\ if and only if the subspace
}$\mathcal{L} =[V \mathcal{H}]\subseteq \mathcal{K}$\textit{\ is
separating for }$ \mathfrak{E}$ \textit{, i.e. for }$Y\in
\mathfrak{E}$ \textit{the relation } $P_{
\mathcal{L}}Y|_{\mathcal{L}}=0$\textit{\ implies }$Y=0. $

Consider the map
\begin{equation*}
\tilde{\Phi}[Y]=V^{\ast }YV,\quad Y\in \mathfrak{E},
\end{equation*}
which is a completely positive map satisfying $\tilde{\Phi}[I]=N,$ which may
be called \textit{complementary} to $\Phi $ (this possibility of
generalizing the notion of complementary map was noticed by Ruskai \cite{rus}
). Then it is easy to see that the above Arveson's criterion of extremality
is equivalent to $\mathrm{Ker}\tilde{\Phi}=0.$ \bigskip

\textbf{Acknowledgments.} The author is grateful to participants of the seminar
``Noncommutative Probability, Statistics and Information'' at the Steklov
Mathematical Institute for useful discussion. He acknowledges partial support of RFBR
grant and of the RAS program ``Mathematical control theory''.

\end{document}